# Credential Control Balance: A Universal Blockchain Account Model Abstract From Bank to Bitcoin, Ethereum External Owned Account and Account Abstraction

**Huifeng Jiao, Dr. Nathapon Udomlertsakul, Dr. Anukul Tamprasirt**

International College of Digital Innovation, Chiang Mai University, Chiang Mai, 50200, Thailand[1]

E-mail: huifeng_jiao@cmu.ac.th, nathapon.u@icdi.cmu.ac.th, anukul@innova.or.th

**ABSTRACT**

Blockchain market value peaked at $3 trillion, fell to $1 trillion, then recovered to $1.5 trillion and is rising again. Blockchain accounts secure most on-chain assets in this huge market (Web-12). This paper initiates **a universal blockchain account model** from a comprehensive review of blockchain account development, encompassing both academic and industry perspectives. This paper uses a **model analysis** method to analysis the account progress and **create** high level **new account model**. And it uses **systematic literature review** method to search, filter, analysis and evaluate the papers about **account models** and analyzes related technology **trade-offs**. Searching with key words: blockchain, account, private key and security in WOS, Scopus and Bitcoin and Ethereum community repositories, this research provides in-depth insights into the design and evaluation of account models, from traditional bank accounts to Bitcoin, EVM-adaptable, and abstraction accounts. Through data-driven **comparisons** of account models (security, cost, adoption), this study also explores future directions and provides an overview of cross-model account theory, guiding further blockchain research. This paper leaves deeper dives into model change drivers, application technology advancements.

## 1    INTRODUCTION

Account is a common-sense concept cause of bank account in your daily life. Richard and Peter's paper lists several bank account types. For example, there are deposit and custody accounts. They rely on three main bank services. These services include payment, insurance, and intermedia (Davies, 2010). Like the early bank account, a blockchain account is a deposit and payment account type now. It will develop into more complicated account types like bank account. This will go with the future of blockchain. Satoshi Nakamoto defined the blockchain and first blockchain account in Bitcoin white paper(Nakamoto, n.d.). Blockchain account hold a digital signature, control bitcoin balance, which provides strong control of ownership.

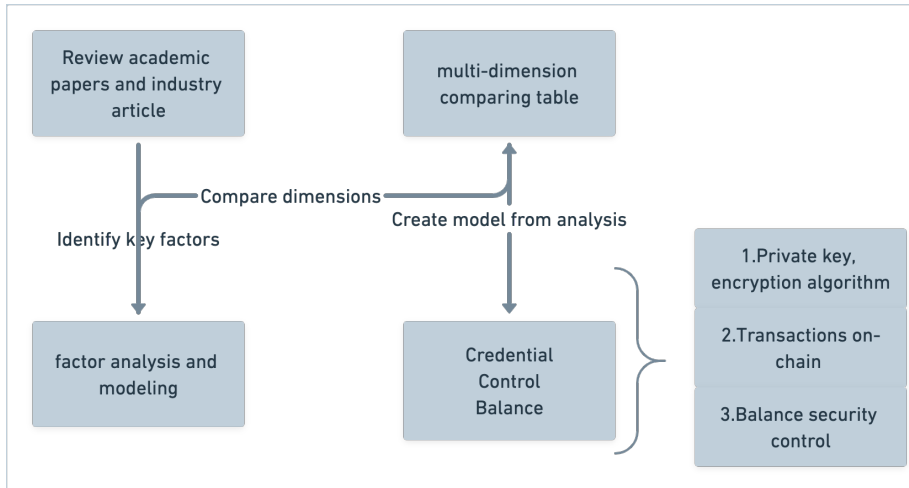

Figure 1: research flow

The crypto market is a rising market, it will become an important market that affect everyone. Speculation and bubbles aside, the crypto market boom also reflects the continued growth of the crypto industry.

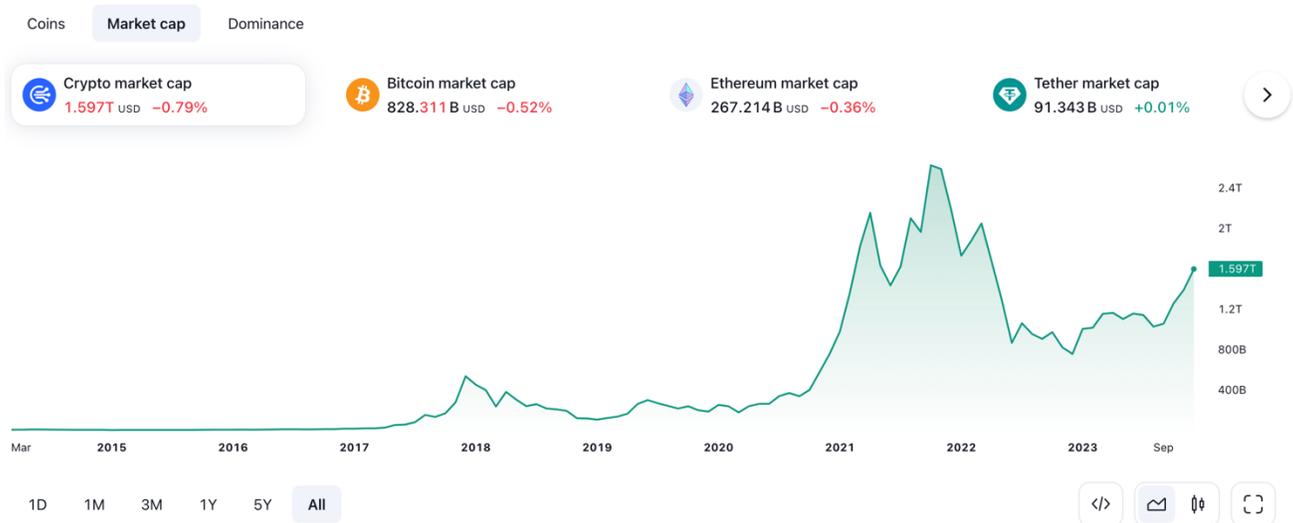

Figure 2: total market value of crypto market at Nov, 2023

Blockchain accounts are evolving. Digital signature(Diffie & Hellman, 2021) and applied algorithm RAS(Rivest et al., 1978), and Merkle hash algorithm(Debnath et al., 2017) technology, and ECDSA(Elliptic Curve Digital Signature Algorithm) (Johnson et al., 2001) and more, are improving the account security level. But some questions remain and wait to implementation (Web-10). These include serious security questions, low TPS (Transaction Per Second) and long verification time. The complex transaction flow, easy-to-lose private key, high technical usage threshold, and high gas cost are also issues. Security, Continence/UX (User Experience,) and Cost, these challenges could hinder widespread human benefit in the digital future.

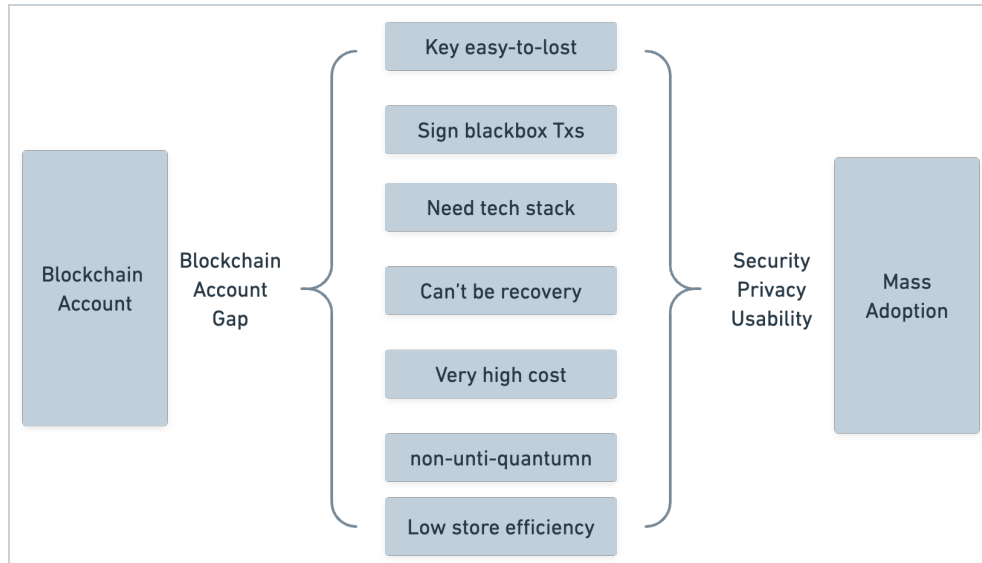

Figure 3: Gap between now account and future

We can set an assumption model with common sense experience: **credentials control balance**. This model is based on an abstract from the bank account. Blockchain account will launch many abilities around these.

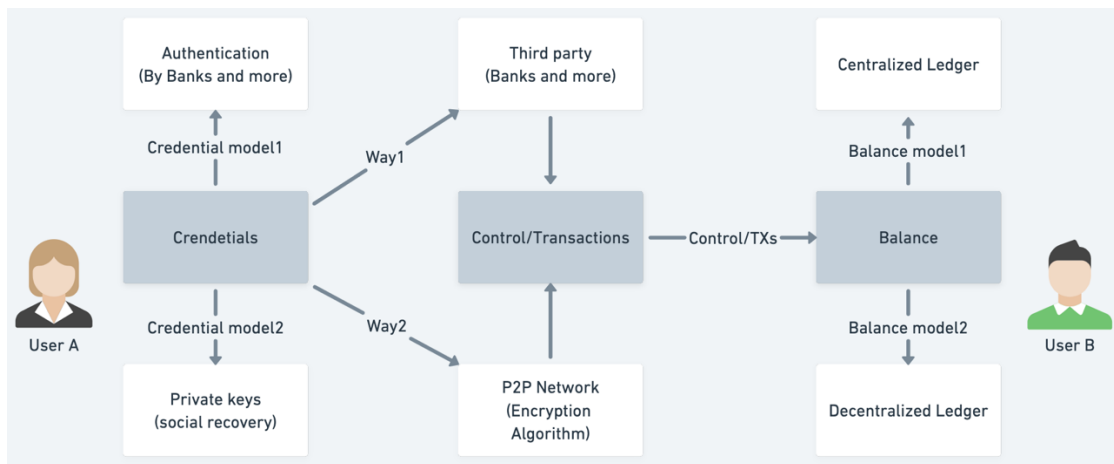

Figure 4: a universal account conception model

The Bitcoin account model (also Blockchain account model) comprises credential, control and balance model. The **credential submodel** is the digital signature based on decentralized consensus. The **control submodel** is hash algorithm and transactions based on a PoW network (Proof of Work) and signature verification methods. The **balance submodel** is UTXO (Unspent Transaction Output) which is like much money with changeable denomination. It is safe and none-double-spending. Bitcoin account model enables P2P transactions "**without the need for a trusted third party**"(Nakamoto, n.d.).

EOA (External Owned Account) is the second type blockchain account in Ethereum (Web-11), but it is easy to lose and be hacked. EOA keeps most assets of the EVM adaptable chains. The EIP55(Web-4) protocol follows a convention for account computation which inherit from BIP39(Web-1) and more. It uses the ECC(Elliptic Curve Cryptography)/Secp256k1(Secure elliptic curve cryptography parameters)(Web-13) (Brown, n.d.) and SHA256(Secure Hash Algorithm)(Web-7) hash encryption method to computation private key to public key which is more security. You must keep your private key carefully. Lost once, lost all.

AA (Account Abstraction) (Web-10) is a developing account type in EVM adaptable chains. It's found in chains like Ethereum, Layer1, Layer2, and more. It empowers mass adoption by breaking the transaction flow into many modules to extend its abilities. The owner address, produced by the private key, is a slot of an on-chain contract. Social recovery can change it. The contract can accept different signature algorithms. The gas payment can be delegated by outer sources. It has many improvements but also has some problems. For example, it has a complex system architecture. It is hard to build and is difficult to keep address consistency. It also has high gas fees. UTXO, EOA, and AA are the prevailing blockchain account models.

Blockchain is a rising industry. There are many public Chains like Ethereum, Optimism, BSC(Binance Smart Chain), Arbitrum and more. Most of the chains follow the pattern of the Bitcoin: using key pairs including private key to sign a digital signature, using public key to verify the signature and transaction execution, which produced by crypto algorithms.

| | Name | Protocols | Addresses | 1d Change | 7d Change | 1m Change | TVL | Stables |
|---|---|---|---|---|---|---|---|---|
| 1 | Ethereum | 1004 | | -0.77% | -0.49% | +11.12% | $32.985b | $71.486b |
| 2 | Tron | 29 | | -0.06% | -4.06% | -6.14% | $7.804b | $52.174b |
| 3 | BSC | 690 | | +0.30% | -3.05% | +8.98% | $3.488b | $32.58m |
| 4 | Arbitrum | 526 | | -1.32% | -0.72% | +7.98% | $2.589b | $2.125b |
| 5 | Solana | 123 | | +1.94% | -3.71% | +33.98% | $1.368b | $1.919b |
| 6 | Polygon | 513 | | +0.55% | -4.27% | -2.65% | $850.44m | $1.31b |
| 7 | Optimism | 218 | | +0.37% | -7.41% | -2.89% | $839.51m | $606.12m |
| 8 | Avalanche | 359 | | -0.16% | -7.76% | -14.43% | $804.13m | $1.104b |
| 9 | Manta | 32 | | -0.32% | +3.63% | +659% | $429.17m | |
| 10 | PulseChain | 34 | | -2.61% | +96.70% | +182% | $412.22m | |
| 11 | Base | 208 | | -0.35% | -6.63% | -10.76% | $401m | $297.29m |
| 12 | Cardano | 33 | | -0.13% | -5.69% | -15.68% | $351.87m | $19.12m |

Figure 5: a total locked value rank in Blockchain on Defilama

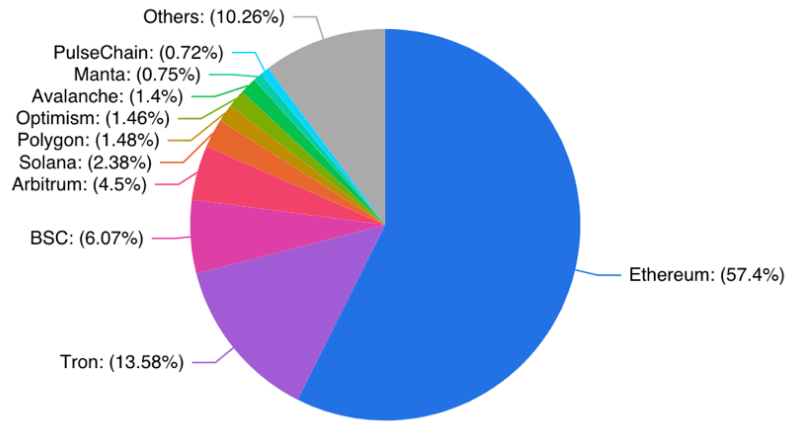

Figure 6 : a pie diagram of market share

This paper used a systematic literature review method. It collects and research most papers about blockchain account. The research purpose is to provide a total review of blockchain account development. It is also to evaluate the **account models**. It compares different account model **trade-offs** using the statistical tables method. Also, we have a model graph to clarify the account's basic actions and interaction flow. We will provide a review of the academic history of a blockchain account.

This research analysis all the structure of the blockchain account model at different times. It launches a new conception, **Universal Blockchain Account: Credential Control Balance**, which means blockchain use hash function and digital signature of Elliptic curve to be a unique and secret credential of your account. And use blocks aggregating all the transactions hashes to guarantee the control of your asset. Use UTXO and Merkle tree and Verkle tree to store and change your balance quickly and security. It will evaluate every model by a triangle: security, convenience, and cost. The models should meet the needs of future mass adoption.

## 2   LITERATURE REVIEW

The conception of an account has evolved over centuries, shifting from simple paper ledgers to modern digital bank accounts, and now venturing into the world of blockchain. This journey signifies a crucial transition: trust and accountability are moving from centralized institutions to individuals empowered by technology.

Traditionally, accounts relied on centralized verification and assurances (think Richard and Peter's accounting(Davies, 2010)). Blockchain technology, pioneered by Satoshi Nakamoto(Nakamoto, n.d.), replaces this with cryptographic proofs and distributed consensus, as explored in works on digital trust and signatures (Debnath et al., 2017). This transformation, fuelled by concerns for privacy and control (highlighted in Bitcoin's white paper (Nakamoto, n.d.) and related research(Debnath et al., 2017)), paves the way for a new era of account management, where users have direct control over their assets while maintaining system integrity.

But how do we ensure secure interactions within a blockchain network? Creating unique and secure blockchain account credentials is key. Cryptographic tools like digital signatures and hash functions form the foundation of account security. Users generate a key pair (private and public) following standards like RFC2459, without relying on central authorities(Housley et al., 1999). The public key, hashed into a unique identifier, becomes the user's blockchain address.

Losing these credentials in a decentralized system can be devastating. Rigorous best practices for recovery and backup are crucial. Literature on digital signatures emphasizes the use of mnemonic phrases ,derived from secure wordlists (Web-2) and Hierarchical Deterministic (HD) wallets(Web-3) to enable recovery from lost private keys. Newer protocols, like social recovery without mnemonics, promise even more robust self-sovereign identity management within blockchain systems.

The core significance of blockchain technology is ability to manage asset balances and facilitate transactions in an environment where there is no centralized trusted entity. Unlike traditional bank accounts, where financial institutions act as trusted intermediaries for balance management and transaction validation, blockchain networks utilize decentralized consensus mechanisms to achieve the same goals. The Bitcoin network's Unspent Transaction Output (UTXO) model is the earliest and most prominent embodiment of this principle, a new paradigm pioneered by Satoshi Nakamoto's seminal white paper(Nakamoto, n.d.). Each UTXO represents a piece of unspent discrete digital currency and is effectively a bearer instrument. Users transact by consuming these UTXOs and generating new UTXOs in chained blocks, thus ensuring that each unit of cryptocurrency is uniquely recorded in the network ledger with a certain capacity (Chakravarty et al., 2020). This mechanism inherently facilitates auditing by establishing a transaction history that does not rely on intermediaries, but on cryptographic verification. Meanwhile, the literature also points to a range of account-based models, including the Ether platform, where individual account balances are updated with each transaction, much like a traditional bank ledger; however, these updates are propagated network-wide by consensus, without the need for a trusted third party (Chakravarty et al., 2020). The blockchain-based mechanism is not only about eliminating centrality in financial operations, but also about providing implicit security and transparency to every participant in the network.

The account infrastructure within blockchain ecosystems can vary substantially, manifesting in differing models of transaction processing and state maintenance. At the core of Bitcoin's approach is the Unspent Transaction Output (UTXO) model, an innovation that meticulously catalogs each fraction of Bitcoin as either spent or unspent, thereby ensuring accuracy in balance computation and resistance to double-spending without necessitating a trusted authority (Nakamoto, n.d.). In contrast, Ethereum embraces an account-based paradigm, primarily through External Owned Accounts (EOAs) and, more recently, Account Abstraction (AA) models. EOAs operate akin to conventional bank accounts with nonce-based transaction ordering, simplifying the state transition logic but paradoxically elevating the risk of key mismanagement and loss (Li et al., 2020). Account Abstraction seeks to ameliorate this by transforming user accounts into smart contracts, permitting more complex access control options, including recovery mechanisms and delegation of transaction fee payments, which are pivotal for user accessibility and security (Web-10). There are some other blockchain account models. Internet Computer(ICP) create a inner-building account model with strong account abilities like instant subaccount with different DApps(Decentralized Application) for high privacy(Web-5).

In the cryptographic domain, account access and security have begun to intertwine with biometric modalities, striving toward enhancing the user experience while simultaneously bolstering security safeguards. The FIDO (Fast Identity Online) protocol epitomizes this expansion, supporting biometric solutions, such as fingerprint recognition, for the verification of blockchain operations, addressing usability without undermining security considerations(Merkle, 1987). A variety of platforms are venturing beyond traditional credentials: Internet Computer Protocol (ICP) leverages user fingerprints for account authentication, thereby intertwining innate biological traits with cryptographic processes, a frontier that is engaging yet not fully charted in terms of privacy implications and inclusivity (Farrugia et al., 2020). And Ethereum community is also evolving the account model based on ERC4337. They try to build a native account abstraction by build-in-client protocols which will improve the efficiency and security, and lower the complexity and cost (Web-6).

Alternatively, Layer2 solutions such as StarkNet and Polygon are exploring scalability and efficiency, signifying an iterative progression towards more refined account models that endorse both functional performance and user assurance (Ma et al., 2021). Some ZK-SNARKs chains try to find a more privacy account model using ZK proof to verify data or transactions without leaking of the original data (Guan et al., 2022). Each model harbours distinct advantages and limitations, mapping a diverse landscape wherein blockchain account infrastructures converge toward a nuanced equilibrium of security, utility, and inclusivity.

A pivotal aspect of blockchain technology is its capacity to sustain a peer-to-peer account balance system without necessitating the intermediation of trusted entities—a stark departure from central authority reliance observed in traditional financial systems. The very bedrock of this decentralized maintenance of account balances is the ingenious application of consensus mechanisms. Such mechanisms foster agreement among network participants on the valid transactions, effectively ensuring the integrity and verifiability of balance updates. Chief among these are protocols like Proof of Work (PoW) and Proof of Stake (PoS), which, despite their divergent operational tactics, serve the unifying goal of distributed trust (Cao et al., 2020). The literature sheds light on these paradigms, such as "Trust and electronic reputation" and more reputation exploring in organizations and individuals are try to build trust on electronic reputation (Web-8) (Zinko et al., 2007) (Web-9).

"An overview of PKI trust models" and more Public Key Infrastructure (PKI)(Buchmann et al., 2013) (S. Khan et al., 2023)(Perlman, 1999), elucidating the evolving notions of digital trust with Certification Authority (CA). But within blockchain network of trust, individual nodes collectively undertake the role traditionally filled by banks or CAs, with technical trust replacing institutional trust. However, the reliance on purely technical consensus has brought forth challenges, notably the scalability trilemma where decentralization, security, and speed engage in a delicate balancing act. So inherited from the mathematics form PKI, blockchain account use the private key and public key system(Koblitz, 1994). The sources dive deep into the intricacies of consensus-driven balance management, offering incisive viewpoints on the current state and envisaging potential enhancements to the digital ledger landscape.

Security considerations are at the core of blockchain account models, as the integrity and trustworthiness of these systems are paramount in their widespread acceptance and success. Over the years, various security breaches and vulnerabilities have been identified, ranging from code exploits in smart contracts to social engineering attacks targeting private keys(Atzei et al., 2017). Notable incidents have underscored the importance of rigorous security measures, fuelling research into the robustness of blockchain systems and the tables of potential attacks. For instance, some studies have focused on detecting illicit activities over blockchain networks, revealing the need for improved monitoring and fraud detection mechanisms (Farrugia et al., 2020). To mitigate such risks, the adoption of advanced encryption techniques, including the Elliptic Curve Digital Signature Algorithm (ECDSA) with threshold signature(Goldfeder et al., 2015) and further refinements like Schnorr signatures, has been proposed. Additionally, multi-signature methods(with threshold signature) have gained traction as they require multiple parties to sign off on transactions, adding an extra layer of security to prevent unauthorized asset transfers (Goldfeder et al., 2014). This methodology diversifies the risk associated with single private key control, a design that has been shown to significantly enhance the security posture of blockchain accounts (Andrychowicz et al., 2016). The continuous evolution of encryption standards and the introduction of multi-signature wallets are emblematic of the persistent efforts to secure blockchain accounts against the varied threat landscape they face.

The privacy importance in blockchain account can't be overemphasized, as it underlies the principles of user autonomy and security. In the blockchain context, the emergence of cryptographic tools such as zero-knowledge proofs and ZK-SNARKs has provided important information for privacy enhancement in blockchain systems (Ma et al., 2021)(Guan et al., 2022) . Zero-knowledge proofs allow the validation of transactions without revealing underlying transaction data, so preserving privacy while maintaining the transaction integrity. ZK-SNARKs—concise non-interactive arguments of knowledge—further streamline these proofs by minimizing computational overhead and interaction between prover and verifier. Fundamental research has established that such methods can be robustly applied in account models to confer privacy on transactional data while still ensuring accountability and verifiability, critical aspects in the blockchain(Ma et al., 2021)(Guan et al., 2022).

And the realization of privacy rights requires trade-offs. The cryptographic protocols are computational intensity, may get challenges to the scalability and latency on blockchain network, potentially impacting transaction throughput and leading to higher costs. The complexity of privacy-preserving techniques may contribute to a higher bar for user comprehension and participation, potentially affecting widespread adoption. Now zero-knowledge techniques provide a compelling privacy ability. It could help to create like a Email account for mass adoption in human digital future.

The blockchain future for mass adoption depends on the blockchain accounts is rely on both technical and human-centric barriers. **Usability** is important; a user's interaction with blockchain technology must be intuitive and seamless to get a widespread acceptance (Brünjes & Gabbay, 2020). Studies have shown that the perceptions and understanding of blockchain among potential users profoundly influence their willingness to adopt the technology (Tsai et al., 2018). Indeed, the complex nature of blockchain can be scary for non-technical users, which can hinder its mainstream using. **Security** is typically at the forefront of user concerns, a robust protective measure that do not compromise usability. **Privacy**, too, is a critical consideration; users must be confident that their transactions and balances remain confidential. Yet, achieving high level security and privacy often comes with an expense of ease-to-use, shaping a trade-offs that researchers and developers must navigate. Designers of blockchain systems are thus challenged to find an equilibrium that addresses these factors, ensuring that the secure and private management of digital assets does not stop potential users through the complexity.

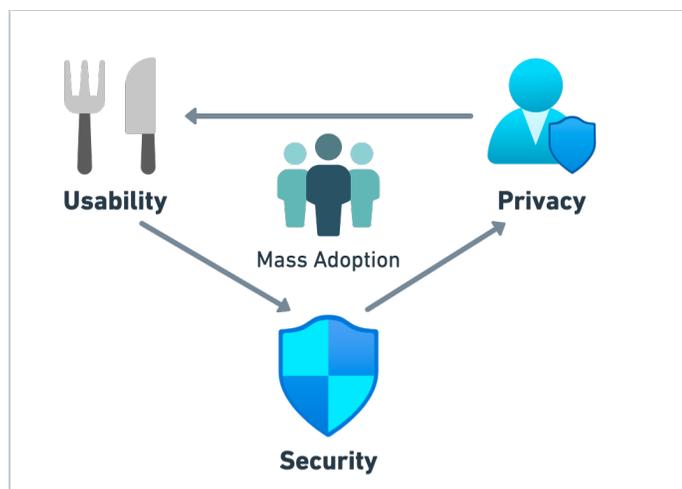

Figure 7: account trade-offss model

Despite the growing volume of research about blockchain, there is a noticeable absence of comprehensive universal model focusing specifically on blockchain account models (Wood & others, 2014). Like an oversight is critical; as the quantity of digital transaction systems expands, it becomes more important to understand the full spectrum of account models—from old bank accounts to practical blockchain variants. The evolution from traditional banking systems to the decentralized architecture offered by blockchains need an in-depth examination of the core components inherent to these models. The trade-offs between usability, security, and privacy have evolved over time. Consequently, an extensive analysis that bridges the gap between foundational banking principles and the innovative paradigms introduced by blockchain technology is not just secured, but crucial. This gap underlines the research community's imperative to develop a universal account model that not only **explain** the technical mechanism along the contiguity of account evolution but also the shifts in paradigms that may influence the course of future **developments** in account management and governance.

The future of blockchain account build on both keeping the immutable core principles about the system trust and meet evolving user expectations. The immutable transactions and the cryptographic blockchain networks remain elementary to users' confidence and the technology's integrity. Recent trends particularly in enhancing user experience and providing regulatory compliance forecast a future where blockchain accounts must balance security with convenience (Atzei et al., 2017). Scalability solutions, whether on-chain advancements or off-chain protocols, continue to evolve to meet the demands of a growing user base(D. Khan et al., 2021). The integration of blockchain technology in various sectors – from finance to healthcare – encourages multidisciplinary research, aiming for a universal account model that simplifies user interaction while maintaining robust security measures. Scholars and industry practitioners are thus tasked with forging pathways towards mass adoption, crafting complicated yet user-friendly interfaces, and ensuring interoperability among diverse blockchain ecosystems (D. Khan et al., 2021). This universal account model aspires to be a blueprint for mass adoption, embodying the trade-offs and synergies between technological capabilities and user needs.

## 3   METHODOLOGY

The methodology of this research is paramount to its efficacy and coherence in addressing the multifarious challenges ingrained in blockchain account technology. **A systematic literature review** methodology is the key point of this study, carefully selected to dissect and navigate the intricate of security, privacy, and obstructions in both **academic** articles and **industry** area.

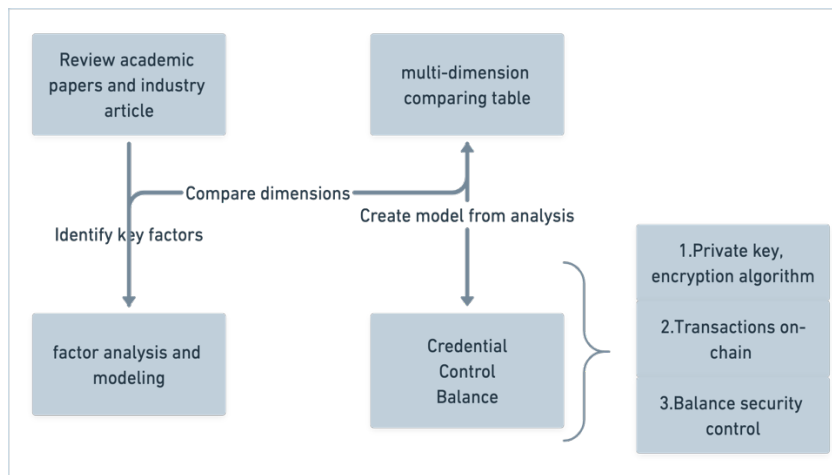

Figure 8: a universal conception account model

This methodology is aligned with the scholarly requisites of comprehensiveness, objectivity, and replicability, which are vital for engendering a robust framework capable of examining the evolution and adoption of blockchain accounts. The agility of this approach allows for the encapsulation of diverse perspectives, engendering a holistic understanding that illuminates both the technical intricacies and the societal implications of blockchain account technology. Beyond mere review, the intent is to construct a nexus of knowledge that can potently inform future innovation, policy-making, and academic inquiry into the digital expanse that blockchain account technology continues to unfurl.

Blockchain account is a practice model in crypto industry, so we use a **model analysis method** to analysis and evaluate the models related to account. There are five phase of industry as the analysis: Bank account, Bitcoin account, EOA account and others, Abstraction Account (Contract Account) and Other EVM Account. It represents the account original source and the development route.

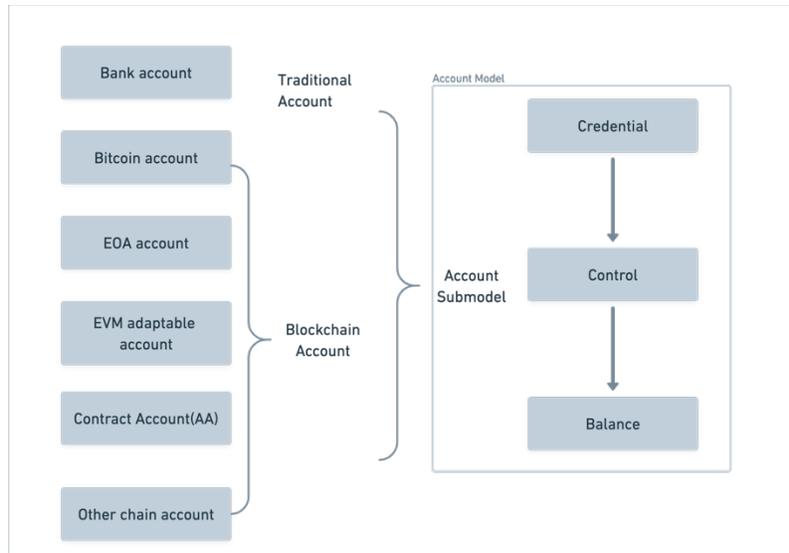

Figure 9: a universal conception account model

This paper split account into **three submodules: credential, control and balance** with the analysis of the account history. It can give a more precise analytical audience and decision-making perspective for follow-up. We use a three-dimension assessment model: security, convenience, and cost which sets privacy as part of the security and splits usability into convenience and cost.

The data collection strategy for this systematic literature review hinged upon a meticulous **aggregation** of industry improvement proposals and scholarly articles that probe the intricacies of blockchain account technology. By considering technical documentation and empirical findings provided by the blockchain industry, particularly white papers and enhancement protocols, this review synthesizes a plethora of data points to form a comprehensive picture of the security, privacy, and usability developments over time. To this end, an **extensive table matrix** was engineered to effectively compare the characteristics, strengths, and vulnerabilities of various blockchain account models. This detailed table not only allows for a clear visualization of comparisons across the research targets but also elucidates the evolution and range of each model, thereby providing an overarching access point for analysis and discussion.

To elucidate the complex interrelations inherent in blockchain account models, our analytical approach leverages a combination of visual and tabular representations. **Diagrams** serve as a foundational tool to visually articulate the intricate relationships between the various components of blockchain accounts—credential, control, and balance—within each model. These schematic illustrations allow for an intuitive comprehension of the nuanced dynamics and structural differences across models. Furthermore, we construct comprehensive tables that facilitate a side-by-side **comparison** of the blockchain account models under study. These tables are instrumental in highlighting discrepancies, delineating **trade-offs**, and contrasting the relative merits and demerits, especially in the realms of security, privacy, and ease of use. Through this **dual-strategy analysis**, we strive to present a clear and accessible synthesis of the data collected, catering to both visual learners and those who prefer detailed, quantitative assessments.

Through this kind of methodology, we would get **muti-dimension review** and give some changing road of the private key making, signature making and verifying, transaction building and block out method, balance modifying and assets moving between multi-accounts. The security tricks in such steps, the convenience of interaction in these complicated flows, the privacy and cost trade-offs on these account model behaviours.

## 4 RESULTS AND DISCUSSION

summary about credential, control, balance, discuss about different model detail on these three sub model. Some references.

Blockchain account secures trillions of crypto assets on the blockchain. But we are **suffering** with the high-tech requirements for opening new blockchain account, private key easy to be lost and many contemporary questions. And the key problem is mass adoption for the future. There will be many questions in security, convenience and cost. We need give a total review on blockchain account and improve it.

Blockchain account is also follow basic account cognition: credential control balance. Blockchain must provide a decentralized credential to remove the trust of third party. Different blockchain account model use different technology to implement decentralization. So "credential control balance" change into "d-credential d-control balance in d-ledger". The first letter "d" means decentralized.

The trade-offs in these models using variant techniques are influenced by a uneven rules. Bitcoin use UTXO balance model to avoid double-spending which is the question discussion mostly in Cryptography Mailing List, Before Bitcoin, B-money is invented by Wei Dai which focus distribution and cryptography security. Bit Gold conception is published by Nick Szabo, which focus on the computing recourse consumption (Popper, 2015).

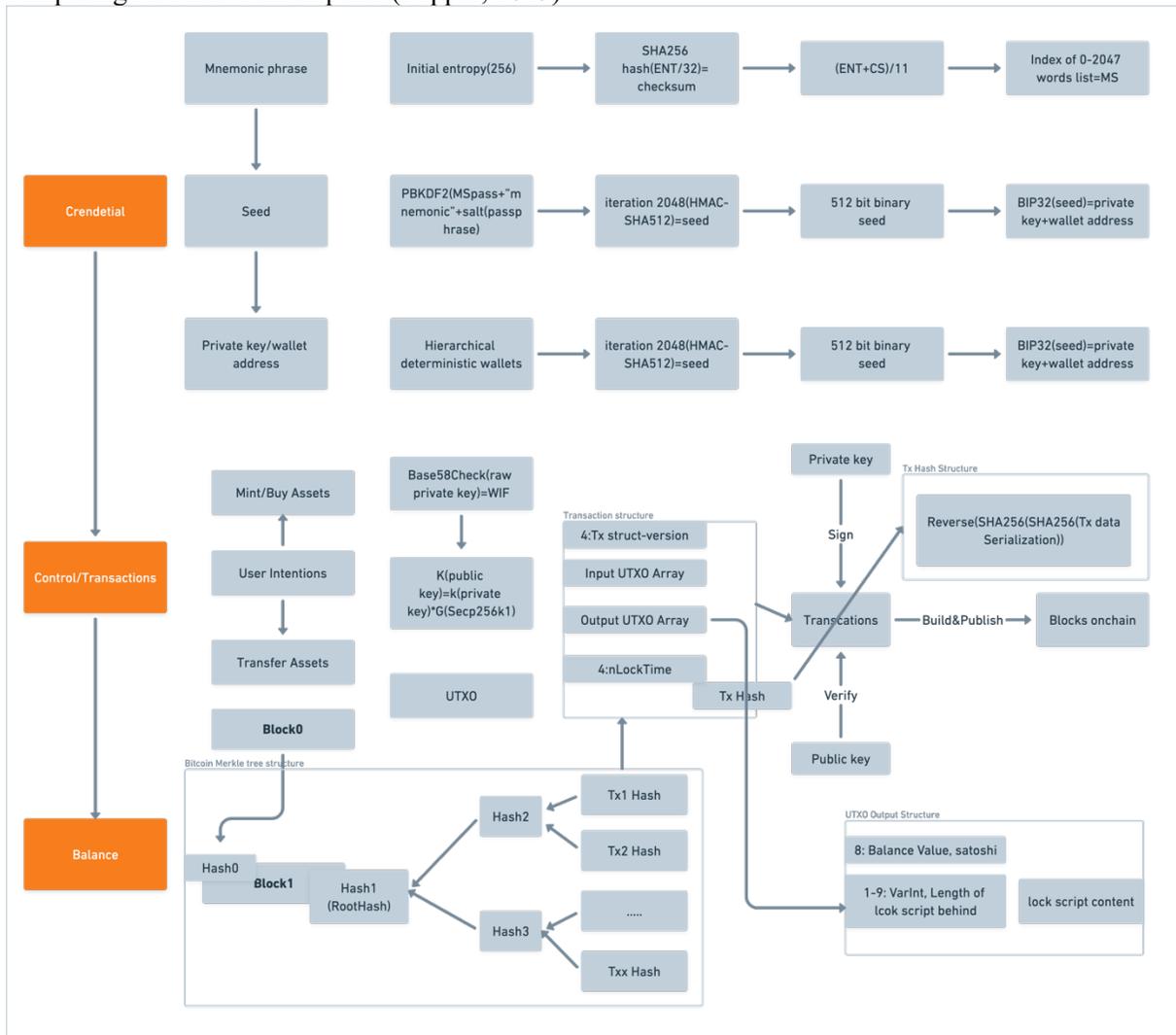

Figure 10: Bitcoin Account Procedure Analysis

The above graph is a detail review of the Bitcoin account mechanism, you can also check the real on-chain blocks and transactions at here (Web-15)(Web-16).

EOA is focus on Engineering convenience, so inherit the credential model, change some algorithm and method, like the control model HD Wallet (Hierarchical Deterministic Wallet), balance model using Merkle tree. And extend the control model to EVM (Ethereum Virtual Machine) with signature.

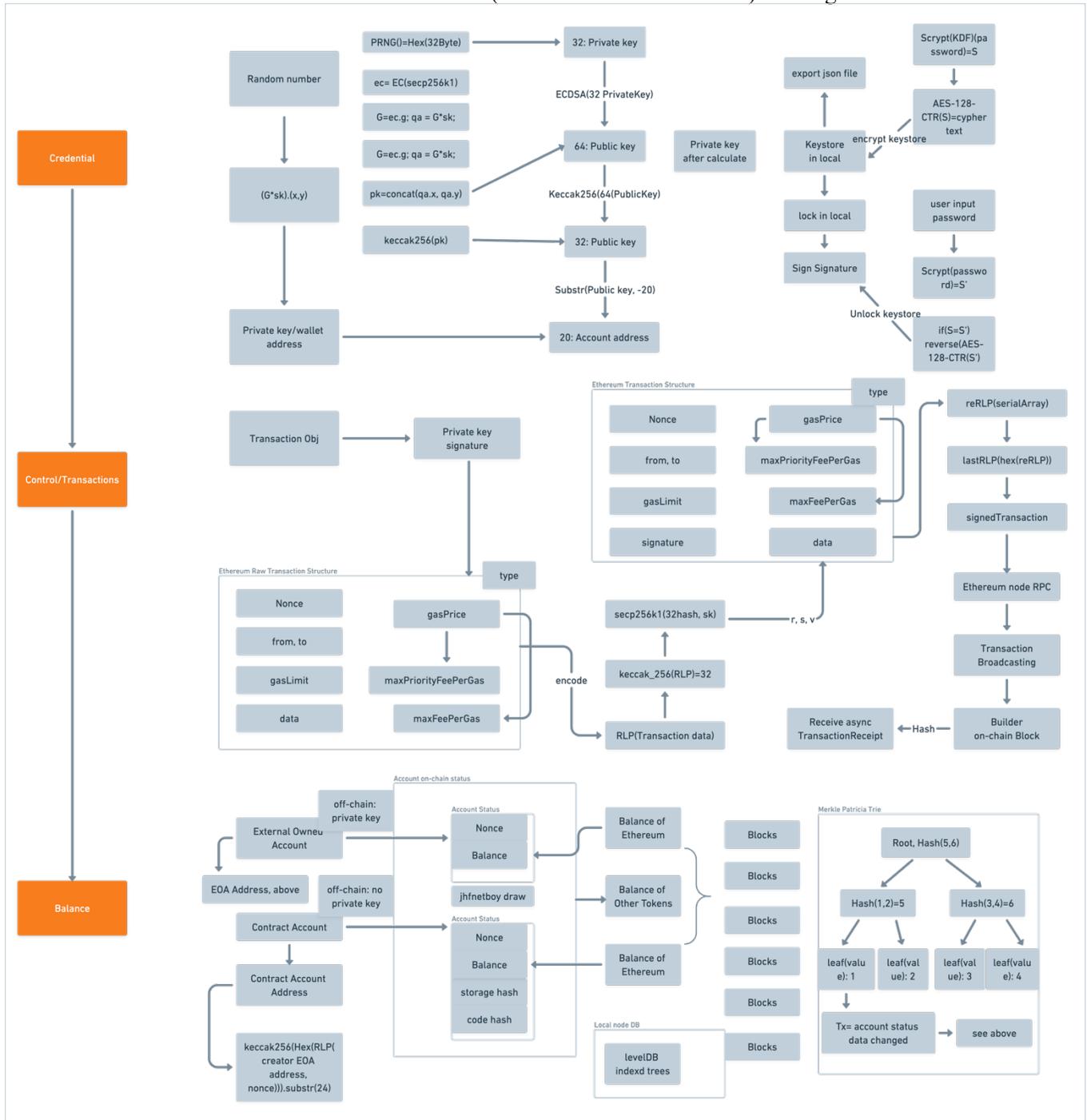

Figure 11: Ethereum Account Procedure Analysis

The above graph is a detail review of the Ethereum account mechanism, you can also check the real on-chain blocks and transactions at here (Web-17) (Web-18) (Web-19).

Account Abstraction focus on the system scalability, not only the balance model, it focus on security like social recovery mechanism, different signature algorithm supporting, Entry point contract monitor and more. It focuses on convenience liken gas fee payment by third party support and programming contract to support session key, Deadman's switch, 2FA with fingerprint, multi-sig with complicate personal or enterprise flow supporting. We can check the simple comparison between the normal EOA transaction and contract account transactions.

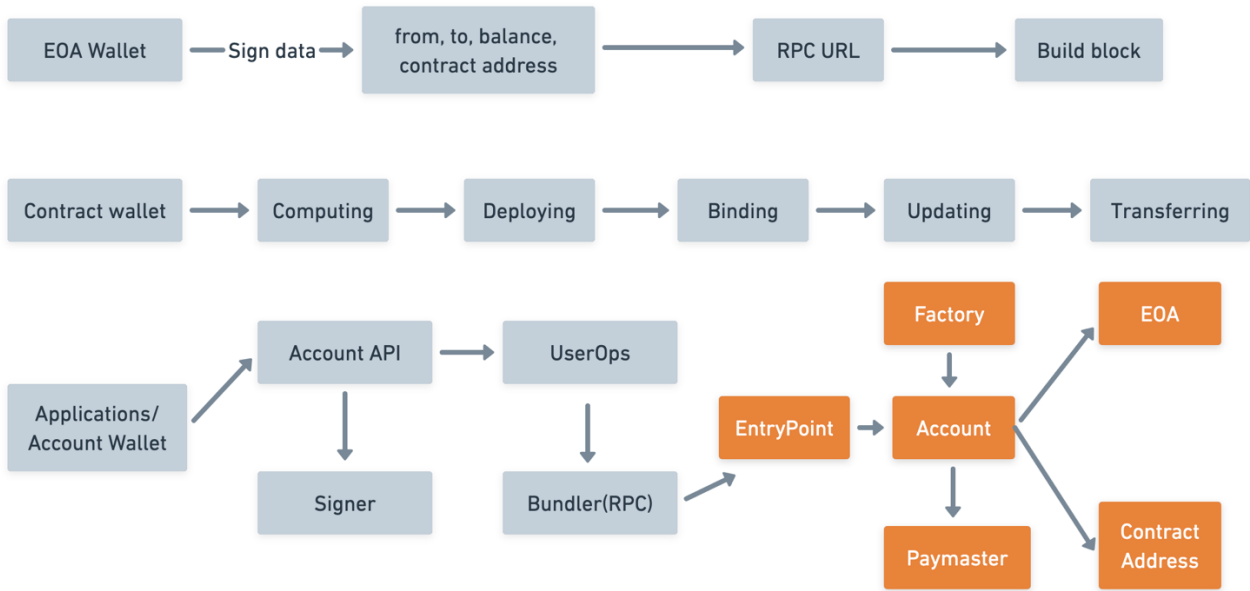

Figure 12: Ethereum Contract Account Procedure Analysis

Account Abstraction, It also need more revised on cost and composability with a balanced evaluation triangle: security, convenience and cost. Or not, the blockchain model of Ethereum can't support the mass adoption of future with billions of users. The industry is trying to launch RIP7560 (Web-6) to build a native Account Abstraction, but we need more ramp up. So, we should dive into this research area to face this challenge.

This research create an array of cutting-edge software tools and techniques in this research. Central to our computational approach was the utilization of Visual Studio Code, an extensible code editor that provided an efficient ecosystem for scripting and testing our analytical algorithms. We augmented this with Code Pilot, an AI-powered coding assistant, that helped streamline the development process by suggesting code snippets and facilitating the debugging of smart contract code. The combination of these tools allowed us to maintain a high level of precision in our analysis. Each step in our data processing and assessment workflow was documented meticulously to facilitate the replication of our research, thereby upholding the integrity and validity of our systemic review. The algorithms and tests implemented were crafted to be transparent and comprehensive, ensuring that our findings and conclusions could faithfully be explored and tested by fellow researchers and industry practitioners.

We select and create blockchain account models with systematic literature review and model analysis, we choose those the most profound impact models on the industry and provide the broadest insights into security, privacy, and adoption area. Consequently, our investigation covers important models such as Bitcoin's UTXO-based system and Ethereum's account-based framework, as these ecosystems represent significant milestones in blockchain technology and are widely recognized within both academic and industry spheres. In addition to these, we considered other emerging and less conventional models that may influence the future trajectory of blockchain account technology. By surveying these varied models, our study encapsulates a diverse demographic, spanning from early adopters embedded in the Bitcoin network to participants in advanced Ethereum smart contract platforms, and further to users of alternative blockchain infrastructures. The relevance of these models to our research aims stems from their substantial contribution to the blockchain landscape, shaping our understanding of how different account structures grapple with the challenges of security, privacy, and mass adoption.

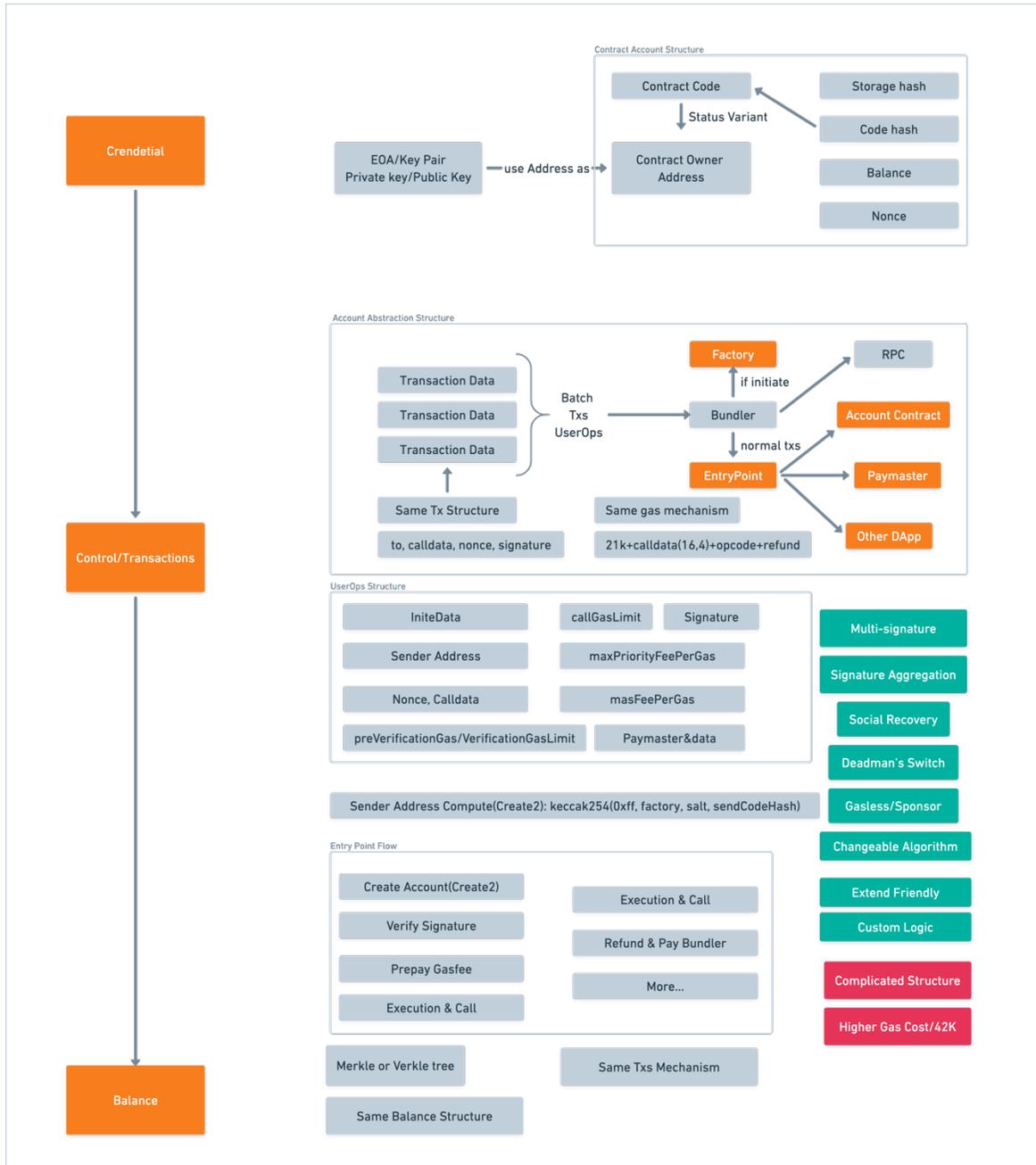

Figure13: Ethereum Account Abstraction Procedure Analysis

Our analytical framework dissected blockchain accounts into three fundamental submodules: credential, control, and balance, to offer a granular perspective on the intricacies of account models. The 'credential' submodule focuses on the authentication mechanisms that ensure rightful ownership, typically through cryptographic key management. 'Control' addresses the mechanisms that govern how transactions are initiated and recorded on the network, including consensus algorithms and smart contracts. Lastly, the 'balance' submodule reflects the methods for maintaining and updating the record of asset ownership. These submodules form the backbone of our comparative model review, interacting synergistically to produce an account's operational profile. To assess these components holistically, we employed a triadic lens of security, convenience, and cost — essential criteria for evaluating the propensity for mass adoption of blockchain technology. This multi-dimensional assessment model illuminates the trade-offs and interdependencies existing within and across blockchain account models, providing insight into their capabilities and limitations from the standpoint of end-user adoption. Through this structured decomposition and assessment approach, our study aims to unravel the intricate fabric of blockchain account security and usability, paving the way for more informed design decisions that could foster wider acceptance.

| Submodule | Behavior | Attributes/Tradeoffs | 1: Bank Account | 2: Bitcoin Account | 3: EOA of Ethereum |
|---|---|---|---|---|---|
| Credential | Create | Who create the key | Centralized Authorities | Mathematics and Encryption Algorithm(MEA) | MEA |
| | | Random number | No | | |
| | Encrypt | How to encrypt | Personal password | PBKDF2(MSpass+"mnemonic"+salt(passphrase)) | PRNG()=Hex(32Byte);ECDSA(32 PrivateKey) |
| | Keep | How to keep | keeping using rights | memonic or bare priate key or json file | keystore |
| | Transaction | Who can launch Tx | User/Authorities | Only User with Private Key(OUPk) | OUPk |
| Control | Encrypt | Txs with encryption | bare database | spec256k1(ECDSA) | spec256k1(ECDSA) |
| | Verify | How to verify | Trust Authorities | Trust enryption and key pair verification | Trust enryption and key pair verification |
| Balance | Logic | Where is the number | Centralized ledger | Decentralized ledger with scattered UTXOs | Decentralized ledger with different account changing Merkle tree |
| | Storage | How to save | RDBS | Blocks include Merkle tree and Txs, Secp256k1, SHA256 | Blocks include Merkle tree and Txs, secp256k1, keccak_256 |
| | Index | Search and interacte | RDBS | Merkel tree | Merkel tree |

Figure 14: Ttrade-offs of account comparison evolution table part1

| Submodule | Behavior | Attributes/Tradeoffs | 4: AA of Ethereum | 5: Other Blockchain Account | 6: Universal Blockchain Account |
|---|---|---|---|---|---|
| Credential | Create | Who create the key | MEA | MEA | MEA |
| | | Random number | | | |
| | Encrypt | How to encrypt | the same with EOA | like keccak and ECDSA | changeable; like keccak and ECDSA |
| | Keep | How to keep | EOA keystore & contract field with social recovery | EVM adaptable are keystore | keystore and social recovery and pubilc guardians |
| | Transaction | Who can launch Tx | OUPk and relation | OUPk | OUPk and trust network |
| Control | Encrypt | Txs with encryption | spec256k1(ECDSA) | spec256k1(ECDSA) and other way | spec256k1(ECDSA) and changeable encryption algorithms |
| | Verify | How to verify | Trust enryption and key pair verification and social relation | Encryption or MPC or mix method | Trust enryption and key pair verification and social relation and DTrust network |
| Balance | Logic | Where is the number | Same with EOA | Like EOA | Like AA with more friendly interface |
| | Storage | How to save | Same with EOA; improve Verkle tree future | Like EOA and other encryption algorithms | Like EOA and AA with changeable encryption algorithms |
| | Index | Search and interacte | Verkle tree | Merkle tree or other index | Merkle tree and future Verkle tree and mult-layer index |

Figure 15: Ttrade-offs of account comparison evolution table part2

The research methodology employed herein is designed to systematically examine the landscape of blockchain account technology, yet it is not without specific scope limitations. Foremost among these are the decision to concentrate on the account models themselves, rather than probing into the multifaceted reasons that underpin the observed trade-offs in security, convenience, and cost. Consequently, while providing a granular analysis of blockchain account types — from their cryptographic foundations to their operational dynamics — this study does not delve into the broader application layer choices such as the strategic utilization of decentralized exchanges (DEXs) and other synthesized account constructs.

Moreover, the account models selected for review, though representative of significant industry milestones, do not encompass an exhaustive inventory of all potential or nascent models. The rationale for such exclusions is manifold, encompassing feasibility within the research time frame, the current stage of industry adoption, and the practical usefulness to the readership in understanding the prevailing state of blockchain account security and operability. These delimitations are explicitly acknowledged to maintain clarity on the intended scope of this review and to signal opportunity scopes for subsequent research endeavours.

## 5 CONCLUSION

This paper reviews the academic articles and industry technology proposals and documents. It create **a universal model** from bank model and Bitcoin account model, Ethereum account and more. It is baed on a total view of blockchain industry account development to guide the future research.

The core conclusion is not only a universal model that explains all blockchain accounts in one model: **Credential Control Balance**. But also includes a universal model with **a tradeoff triangle model** to keep evolution balance. It is a **triangle of security, privacy, usability** to future mass-adoption.

And this research also get a series of **comparison tables**, which lead the direction and **regulate** the **trade-offs** of the model evolution gains and losses. This table contain items about the **basic behaviors and attributes** of every sub model. The most basic **decisions** we have to measure and choose from are based on different variations of these attributes and behaviors.

This paper finds that, blockchain account comprises **three sub models: credential, control and balance**. From Bitcoin to EOA, to AA and Native AA, blockchain account try to improve their security and convenience. Although there are many questions in security, convenience and cost for future mass adoption. It is also simple **description** of account model: credential control balance.

But it important in overall **consideration of three elements: security, privacy and usability**. Trillions of assets are secured by blockchain accounts, billions of users will step into these human digital public goods. This research points out the that these trade-offs of convenience and cost are complicated and be caution of the decentralization of those improvements of blockchain account.

We think the roadmap of Account Abstraction is right. The **challenge** is, that the two shortcomings of improving **convenience** and lowering the **cost** of use, while simultaneously improving the resulting security reduction in parallel. Otherwise, it won't be possible to match future large-scale applications.

It is possible to use this universal blockchain account model to analysis and **improve** the reality account model. We use the **comparison table** to evolute the AirAccount project model, balance the security, privacy and usability.

Let us just exploring the blockchain future like Ninja Turtles with a cooperative mindset (Web-14).

## 6 ACKNOWLEDGEMENTS

The authors would like to thank the following people for their contributions to this work:
Dr. Anukul, my advisor, for his encouragement and support on my project AirAccount.
Dr Nathapon, my co-advisor, for his guidance and support throughout the paper and research.
4Seas Community support, for their grant and kindly support on holding workshop and hackathon.
Our classmates FengHan, Lily and Miko, YT for their helpful suggestions and feedback.
My families, for their love and encouragement.